\documentclass[12pt]{article}
\usepackage{epsfig}

\newcommand{\bea}{\begin{eqnarray}}
\newcommand{\beq}{\begin{equation}}
\newcommand{\eea}{\end{eqnarray}}
\newcommand{\eeq}{\end{equation}}

\newcommand{\nn}{\nonumber}

\newcommand{\lsim}{\raise0.3ex\hbox{$\;<$\kern-0.75em\raise-1.1ex\hbox{$\sim\;$}}}
\newcommand{\gsim}{\raise0.3ex\hbox{$\;>$\kern-0.75em\raise-1.1ex\hbox{$\sim\;$}}}
\newcommand{\bpm}{\begin{pmatrix}}
\newcommand{\epm}{\end{pmatrix}}
\newcommand{\eq}[1]{Eq.~(\ref{#1})}
\newcommand{\unity}{{\hbox{1\kern-.8mm l}}}

\newcommand{\rfn}[1]{(\ref{#1})}

\begin{document}

\pagestyle{empty}
\begin{flushright}
CERN-PH-TH/2008-140
\end{flushright}
\vspace*{5mm}
\begin{center}
{\large {\bf Supersymmetric Grand Unification and
Lepton Universality in $K\rightarrow \ell\nu$ Decays
}} \\
\vspace*{1cm}
{\bf John Ellis~$^1$, Smaragda Lola~$^2$} and {\bf Martti Raidal~$^3$} \\
\vspace{0.3cm}
$^1$ Theory Division, Physics Department, CERN 1211, Geneva 23, Switzerland \\
$^2$ Department of Physics, University of Patras, GR-26500 Patras, Greece \\
$^3$ National Institute of Chemical Physics and Biophysics \\
Ravala 10, Tallinn 10143, Estonia

\vspace*{2cm}
{\bf Abstract}
\end{center}
\vspace*{2mm}

Motivated by the prospects for an improved test of lepton universality
in $K\rightarrow \ell\nu$ decays by the NA62 experiment at CERN, we study predictions
for the possible lepton non-universality in $K\rightarrow \ell\nu$ decays in supersymmetric models.
Violations of $\mu-e$ universality in this process may originate from mixing
effects in the right-handed slepton sector, providing a unique window into this aspect of
supersymmetric flavour physics in the large-$\tan\beta$ region.
Minimal unification scenarios with universal soft
supersymmetry-breaking terms at the GUT scale would predict
negligible violation of lepton universality. However,
lepton non-universality
may be observable in non-minimal grand unified models with
higher-dimensional terms contributing to fermion masses, in which case
renormalization effects above the GUT scale may
enhance the mixing among the right-handed sleptons.
This could leads to observable lepton non-universality
in $K\rightarrow \ell\nu$ decays
in specific regions of the parameter space with high $\tan \beta$,
large $A$ terms and small charged Higgs boson mass. Observable non-universality in
$K\rightarrow \ell\nu$ decays would be correlated with a large value of
$BR(\tau\rightarrow e \gamma)$. The experimental upper limit on the electric dipole
moment of the electron could be reconciled with leptogenesis,
if the latter occurs at a relatively low scale, which would also
alleviate the cosmological gravitino problem. Even if lepton non-universality
is not seen in the near future, one may nevertheless obtain significant
constraints on the model parameters and unknown
aspects of right-handed fermion and sfermion mixing.

\vspace*{1cm}
\begin{flushleft}
CERN-PH-TH/2008-140 \\
September 2008
\end{flushleft}

%\vfill\eject

\setcounter{page}{1}
\pagestyle{plain}

\pagebreak

\section{Introduction}

A large number of  experiments have measured
neutrino oscillations~\cite{S-Kam, SNO, KamLAND, K2K, MINOS},
thereby providing important information on the neutrino mass differences 
and mixing angles~\cite{N-data}. 
Within the framework of supersymmetry, 
massive neutrinos lead to charged-lepton-flavour violation (LFV) via radiative corrections to 
sfermion masses~\cite{BM,KO}, that may be observable in forthcoming
experiments. The predictions of models of massive neutrinos for
processes such as $\mu \rightarrow e \gamma$, $\mu \rightarrow 3 e$,
$\tau \rightarrow \mu \gamma$, $\mu \to e$
conversion on heavy nuclei and sparticle decays at the LHC~\cite{LFVlhc}
have been studied extensively~\cite{KO}. These predictions are
frequently very close to the current experimental limits~\cite{KO,lfv-neu,Casas:2001sr,HisNo}, 
and may be further
refined by requiring successful leptogenesis~\cite{LEP-CP-NU,Ellis:2002xg}
and sneutrino inflation \cite{sn-inflation}.

In addition to charged-lepton decays, rare decays of mesons 
are also of potential interest. In a previous work, we studied in detail rare kaon
decays~\cite{Kaons-John} to $\mu e$ pairs with and without accompanying pions,
finding that radiative corrections related to neutrino mixing may induce significant
rates, even when starting from universal initial conditions for the soft terms at a
high-energy input scale.
In these examples, as well as in most low-energy LFV processes,
the relevant mixing arose dominantly from {\em left-handed}
slepton mixing induced via the renormalization-group equations (RGEs). 

It has recently been pointed out that mixing effects in the {\it right-handed} sfermion sector
can be probed very sensitively by checks on $\mu - e$ universality in the decays
\[
K\rightarrow \ell\nu \; , \; \ell \equiv e, \mu
\]
which can be generated by flavour non-universality in an
effective $\bar{\nu}_\tau \ell_R H^{+}$ coupling~\cite{Masiero,Brignole:2004ah,Paradisi:2005tk}.
In general, the uses of meson decays as probes of physics beyond the Standard
Model (SM) are complicated by hadronic uncertainties.
However,  working with the ratios 
of the electronic and muonic decay modes, in this case
$R_{K}\! \equiv \!\Gamma(K\!\rightarrow \!e\nu)/\Gamma(K\!\rightarrow\! \mu\nu)$,
the hadronic uncertainties cancel to a large extent, allowing a precise confrontation
between theory and experiment. The current bound on $R_K$ is given by ~\cite{RKex}:
\bea
R^{exp}_{K} = (2.457 \pm 0.032) \cdot 10^{-5},
\eea
which is to be compared with
the SM prediction $R^{SM}_{K}=(2.472\pm 0.001)\cdot 10^{-5}$~\cite{RK-SM}.
The NA62 experiment at CERN now plans a significant improvement in the experimental 
accuracy,  expecting to reduce the uncertainty in $R_K$ to $\pm 0.003$.

Any violation of $\mu - e$ universality in $K\rightarrow \ell\nu$ decays would constitute
unambiguous evidence for new physics. In particular, within a supersymmetric
framework, it would provide crucial information on right-handed slepton mixing, thereby
complementing in an important way the other LFV processes studied previously~\cite{Masiero}.
As we discuss later, in schemes with universal soft scalar masses
at the GUT scale, the experimental bounds on other LFV processes imply that
RGE effects below the GUT scale would be insufficient to generate 
non-negligible $\mu - e$ non-universality in $K\rightarrow \ell\nu$ decays.
However, right-handed slepton mixing and 
$\mu - e$ non-universality might arise
through RGE effects above the GUT scale~\cite{Barbieri:1994pv}
in models where universality of the soft
supersymmetry-breaking contributions to the right-handed slepton masses is
assumed at some higher input `gravity' scale $M_{\rm grav}$.
This might then lead to observable non-universality in $K\rightarrow \ell\nu$
if $\tan \beta$ is large and other conditions on the supersymmetric model parameters
are also met. However, the simultaneous presence of both left- and right-slepton flavour mixings, 
together with very large values of $\tan\beta,$ would in general imply too large rates  
for the LFV decays end electric dipole moments (EDMs) of charged leptons.
Therefore, in order for non-universality to be observable in $K\rightarrow \ell\nu$,
consistency with the present bounds on LFV imposes non-trivial conditions
on the flavour physics  as well as the supersymmetry-breaking pattern.

In this paper we study the patterns of the soft supersymmetry breaking
terms  required for obtaining observable  renormalization induced 
$\mu-e$ non-universality effects in
the NA62 experiment at CERN.
As an initial condition we assume the SUSY breaking parameters to be flavour 
universal at $M_{\rm grav} > M_{GUT}$ and consider the RGE running 
of the soft supersymmetry-breaking mass parameters both above and below the GUT scale. 
We assume the seesaw mechanism~\cite{seesaw} with three singlet neutrinos, and 
we use the observed neutrino masses and mixing angles as inputs. We apply a 
parameterization via a Hermitian matrix $H$~\cite{Parametrisation} employing the orthogonal
parameterization~\cite{Casas:2001sr}
to calculate the corresponding
singlet neutrino Yukawa couplings  $Y_\nu$ and masses $M_N.$
This parameterization greatly facilitates keeping the RGE-induced left-handed
slepton flavour structure under control. 
Within the minimal supersymmetric SU(5) GUT, the flavour mixing of the right-handed sfermions
is RGE-induced above the GUT scale by the Cabibbo-Kobayashi-Maskawa matrix. 
%In our scenario, together with heavy neutrino RGEs, this would lead to too large rates
%of LFV. 
However, it is well known that this minimal SU(5) GUT relates
$m_e$ and $m_\mu$ incorrectly to $m_d$ and $m_s$. This defect can be
cured by adding supplementary terms in the $d$-quark and charged-lepton 
mass matrices and in the coloured triplet Higgs Yukawa couplings 
originating from higher-order, non-renormalizable terms in the
effective superpotential below $M_{\rm grav}$~\cite{his2}. The corrected Yukawa couplings 
leave their imprint on the flavour structure of the right-slepton supersymmetry 
breaking parameters via renormalization above the GUT scale.

Throughout our analysis, 
we require the magnitude and pattern of supersymmetry breaking parameters
to be consistent with supersymmetric Dark Matter, the  baryon asymmetry of the 
Universe, and with all present bounds on flavour-violating decays and EDMs.  
This imposes nontrivial requirements on the pattern of SUSY breaking parameters.
As an example,
one pattern of  supersymmetry breaking  which simultaneously gives the 
desired Higgs and sparticle mass spectrum, the correct amount of DM,
and large RGE induced non-universality effects,
is a tuned version of the so called Higgs boson
exempt no-scale supersymmetry breaking \cite{noscH}. 
This scenario allows us to generate small charged  
Higgs boson masses, while keeping all other soft mass terms heavy so that all other relevant
observables like $(g-2)_\mu,$ $B\to \mu\mu, $ $b\to s \gamma$ etc. are consistent with the measurements.
One consequence of this sample SUSY breaking point is that
supersymmetric particles would be difficult to discover at the LHC,
whereas the charged Higgs boson
should be relatively easily accessible at the LHC~\cite{Ball:2007zza}.

Within this framework, we
we find  examples with values of the renormalization induced non-universality parameter
$\Delta R_K$ as high as $\sim {\cal O}(10^{-2})$ to $(10^{-3})$,  well within the reach of the NA62 experiment. However, in order to achieve this,
a very constrained flavour structure for Yukawa matrices is required in order to
keep LFV decays under control while generating large non-universality effects
in $K\rightarrow \ell\nu.$ 
As a result, we find a strong correlation between the decay $\tau\to e\gamma$ and the
size of $R_K.$ Observation of one of them would, knowing the SUSY parameters
and the mass of the charged Higgs boson,  
predict the other.  In order to have observable  $R_K$, the charged 
Higgs boson must be light while the other supersymmetric particle masses must be 
heavy in order to suppress $\tau\to e\gamma$ below the experimental bounds.

In this scenario large EDMs of charged leptons are induced if there are phases in the
complex neutrino Yukawa couplings $Y_\nu$ 
as would be needed to generate the baryon asymmetry 
of the Universe via leptogenesis~\cite{Fukugita:1986hr}. 
At large $\tan\beta,$ and due to the simultaneous presence
of both $\delta \tilde m_{LL}$ and  $\delta \tilde m_{RR}$ flavour mixings, 
neutrino Yukawa-induced
contributions to the $i$-th lepton EDM is strongly enhanced due to 
 the dominant term $(\delta \tilde m_{LL}^2 \;m_\ell \mu\tan\beta\;  \delta \tilde m_{RR}^2  )_{ii}.$  
 While in the minimal SUSY seesaw models the induced charged lepton EDMs are, in the most
 optimistic case, a few orders of magnitude below the present experimental bound 
 \cite{JM,Masina:2003wt,Farzan:2004qu},
 in our scenario the EDMs can be significantly larger.
 Suppressing the electron EDM below the experimental bound 
 $d_e<1.6\cdot 10^{-27}$~e~cm~\cite{Regan:2002ta} by assuming
small phases in the neutrino Yukawa couplings is possible but, in the absence of 
a concrete theory of phases, may be unnatural, and would also
suppress the  CP asymmetry for leptogenesis. We find that the natural way to
suppress the electron EDM in this scenario is 
related to the flavour structure of the heavy neutrino Yukawa couplings and,
consequently, the flavour structure of the induced soft SUSY breaking terms.
Using the $H$ parameterization of neutrino Yukawa couplings~\cite{Parametrisation}, 
assuming $H_{11}\ll H_{22,33}$ would imply $M_{N_1}\ll M_{N_{2,3}}$ and, therefore,
small Yukawa couplings of the lightest heavy neutrino $N_1.$ Hence, for a given $M_{N_1}$
we find an {\it upper limit} on the electron EDM. 
We argue that the gravitino problem~\cite{GRprob} in
supersymmetric theories, which sets upper limits on the reheating temperature of the Universe
and therefore  requires a relatively light $N_1$ for successful leptogenesis,  also
provides a solution to the EDM problem. For $M_{N_1} \sim 10^{8}\; (10^{6})$ GeV the electron EDM 
is bounded as $d_e  \lsim 10^{-28}\;(10^{-30})$~e~cm which is within the reach of the
proposed electron EDM experiments~\cite{demille}.

The outline of our paper is the following. In Section 2 we review the details of 
non-universality effects in  the decays $K\to \ell \nu$.  In Section 3 we discuss the 
RGE effects in supersymmetric models, including running both below and above the GUT scale.
Numerical examples are given in Section 4, and we conclude in Section 5.

\section{Sfermion Mixing, $\mu-e$ Universality in $K\rightarrow \ell\nu$ Decays
and other Observables}

Probing charged-lepton universality
in $K\rightarrow \ell\nu$ decays ~\cite{RK-SM} is interesting in view of the very promising
experimental prospects and  since, in a supersymmetric framework,
this decay probes mass mixing between the right-handed charged sleptons, $\tilde{m}_{RR}$. 
In contrast to mixing between the left-handed charged sleptons, $\tilde{m}_{LL}$, rare decays 
and other processes have given us relatively little information yet about $\tilde{m}_{RR}$.
It is clear~\cite{Masiero} that there would be significant sfermion mixing in the
presence of general non-universal soft masses at the GUT scale. Here, however,
we will focus on lepton-flavour violation (LFV) induced by the renormalization-group equations
for soft terms, in particular through the effects of right-handed neutrinos below the GUT scale
and through RG evolution above the GUT scale in models of grand unification. 

The decay $K\rightarrow \ell\nu$ has been discussed in detail in~\cite{Masiero,Brignole:2004ah,Paradisi:2005tk},
where its magnitude was shown to be 
dominated by~\cite{Masiero}
\bea
\label{lfclfv}
R^{LFV}_{K}\simeq R^{SM}_{K}
&\bigg[&\bigg|1\!-\!\frac{m^{2}_{K}}{M^{2}_{H}}
\frac{m_{\tau}}{m_{e}}\Delta^{11}_{RL}\,\tan^{\!3}\!\beta\bigg|^{2} +
\bigg(\frac{m^{4}_{K}}{M^{4}_{H}}\bigg)
\!\bigg(\frac{m^{2}_{\tau}}{m^{2}_{e}}\bigg)
|\Delta^{31}_{R}|^2\,\tan^{\!6}\!\beta
\bigg],
\label{master}
\eea
where 
\beq
\Delta^{\ell \ell}_{RL}\! \simeq \! -\frac{\alpha_{1}}
{4\pi}\mu M_1 m^{2}_{L} m^{2}_{R} 
\delta^{\ell 3}_{RR} \delta^{\ell 3}_{LL}
I^{''} (M^{2}_{1},m_L^2,m^{2}_{R})),
\label{DelRL}
\eeq
and 
\beq
\Delta^{3\ell}_{R}\! \simeq \! \frac{\alpha_{1}}
{4\pi}\mu M_1 m^{2}_{R} \delta^{3\ell}_{RR}
\left[I^{'}\!(M^{2}_{1},\mu^2,m^{2}_{R})\!-\!(\mu\!\leftrightarrow\! m_{L})
\right].
\label{DelR}
\eeq
In these expressions
${I}$ is the standard three-point one-loop integral
\beq
{I}(x,y,z) \equiv \frac{ xy \log\frac{x}{y}  +yz \log\frac{y}{z} 
+ zx \log\frac{z}{x}}{ (x-y) (z-y)(z-x)},
\eeq
and $I^{'}(x,y,z) \equiv \frac{dI(x,y,z)}{dz},$  $I^{''}(x,y,z) \equiv \frac{d^2 I(x,y,z)}{dy dz}.$
As usual we denote
\beq
\delta^{ij}_{XX}
\! \equiv \!({\tilde m}^2_{\ell})^{ij}_{XX}/m^{2}_{X} \; \; (X=L,R),
\eeq
and for the rest of the paper we will drop the flavour indices
in $\delta^{ij}_{XX}.$
The first term in (\ref{lfclfv}) features a double insertion of LFV mixing,
and interferes with the SM contribution, whereas the second term
clearly has no such interference.
Note that we neglect a term proportional to $\Delta^{32}_{R}$, which
is suppressed by a factor $m^{2}_{e}/m^{2}_{\mu}$ with respect to the term
proportional to $\Delta^{31}_{R}$.
Similarly, we neglect the contributions from left-slepton mixing $\Delta_L$
as those are numerically subleading~\cite{Paradisi:2005tk}.
In our numerical calculations we use the full expressions from~\cite{Brignole:2004ah}
rather than just the dominant terms (\ref{master}, \ref{DelRL}, \ref{DelR}).
However, the latter expressions are sufficient for discussion of the
new physics non-universality effects in kaon decays. 

The dependence of the deviation from universality in the $K \to \ell \nu$
decay rates on $\Delta^{11}_{RL}$ and $\Delta^{31}_{R}$
is not complicated;
it is clear from the formulae (\ref{master}, \ref{DelRL}, \ref{DelR})
that larger rates are expected for large $\tan\beta$,
a light `heavy' Higgs mass $M_H$, large $\mu$ (note, in particular,
that the dominant $\Delta_{RL}^{\ell \ell}$ contribution
is proportional to $\mu$), and small slepton masses
(in order to avoid suppressions in the three-point loop functions; this can be true
for the right-handed staus, in particular).
Specifically, for $M_H = 180$~GeV and $\tan\beta = 50$, 
one obtains
\bea
\delta R^{LFV}_{K} \simeq 
10^7 
[ (\Delta^{31}_R)^2 + (\Delta^{11}_{RL})^2 
- 0.0006  \Delta^{11}_{RL} ] .
\eea

In general: \\
(i) For the range of parameters where $\delta_{LL,RR}$ 
have small and comparable
magnitudes, the
interference term  proportional to 
$\Delta^{11}_{RL}$ would be expected to dominate over 
$(\Delta^{31}_R)^2$ and $(\Delta^{11}_{RL})^2$. \\
(ii) In the case that $\delta_{LL} \ll \delta_{RR}$, 
$\delta R^{LFV}_{K}$ scales as the $(\Delta^{31}_R)^2$ and thus   $(\delta_{RR})^2$ terms. \\
(iii) For larger $\delta_{RR,LL}$,
both quadratic and linear terms may be important in $R_K$.  

Barring a cancellation, an experimental measurement with an error $\Delta R_K \sim 0.003$
would provide sensitivity to $\Delta^{11}_{RL} \sim 5 \times 10^{-7}$ 
with  a significantly smaller $(\Delta^{31}_{R})^{2}.$
On the other hand, for a very small $\delta_{LL}$ and thus $\Delta^{11}_{RL}$,
the sensitivity to $\Delta R_K \sim 0.003$ is compatible with 
$\Delta^{31}_{R} \sim 1.7 \times 10^{-5}$ (which for the above quoted  
optimal  set of supersymmetric
parameters would correspond to $\delta_{RR} \sim 0.12$).

Because of the prefactors in $\Delta^{3\ell}_{R}$, 
unless $x=y=z$ to a great accuracy (which is not expected,
in view of RGE effects), one would typically expect
$\Delta^{3\ell}_{R} \leq 10^{-3}$ even in models with enhanced non-universalities, such 
as in \cite{andrea}.
Values of this order of magnitude are potentially interesting for experiment. 
However, if non-diagonal scalar terms are induced only by
RGE effects, one expects rather smaller values of the $\delta^{ij}_{XX}$
(and hence $\Delta^{11}_{RL}$ and $\Delta^{31}_{R}$) than in models
where universality is explicitly violated.
In typical scenarios, one expects the RGE-generated 
right-handed mixings to be small, whereas  the
$\delta_{LL}$ are found to be generically larger.
Nevertheless, it is clear that if there is a signal 
in $K\rightarrow \ell\nu$ decays in the near future, this
would imply a non-negligible  right-handed 
slepton mixing, and would  inevitably lead to very constrained
scenarios, particularly for models with universal 
initial conditions for the soft terms:
models with large $\tan\beta$, light right-handed staus 
and large A-terms  would be favored.

Moreover, since  observable non-universality effects in  $K$ decays  
would require non-negligible $\tau -e$ mixing in the
RR slepton sector, the LFV decays $\tau\to e\gamma$ must 
inevitably be large, and correlated with the non-universality.
This among others would imply strong constraints on the latter from the bound 
$BR(\tau \to e \gamma) < 1.1 \times 10^{-7}$ \cite{tauegamma}.

\begin{figure}[!h]
\includegraphics*[height=7.5cm]{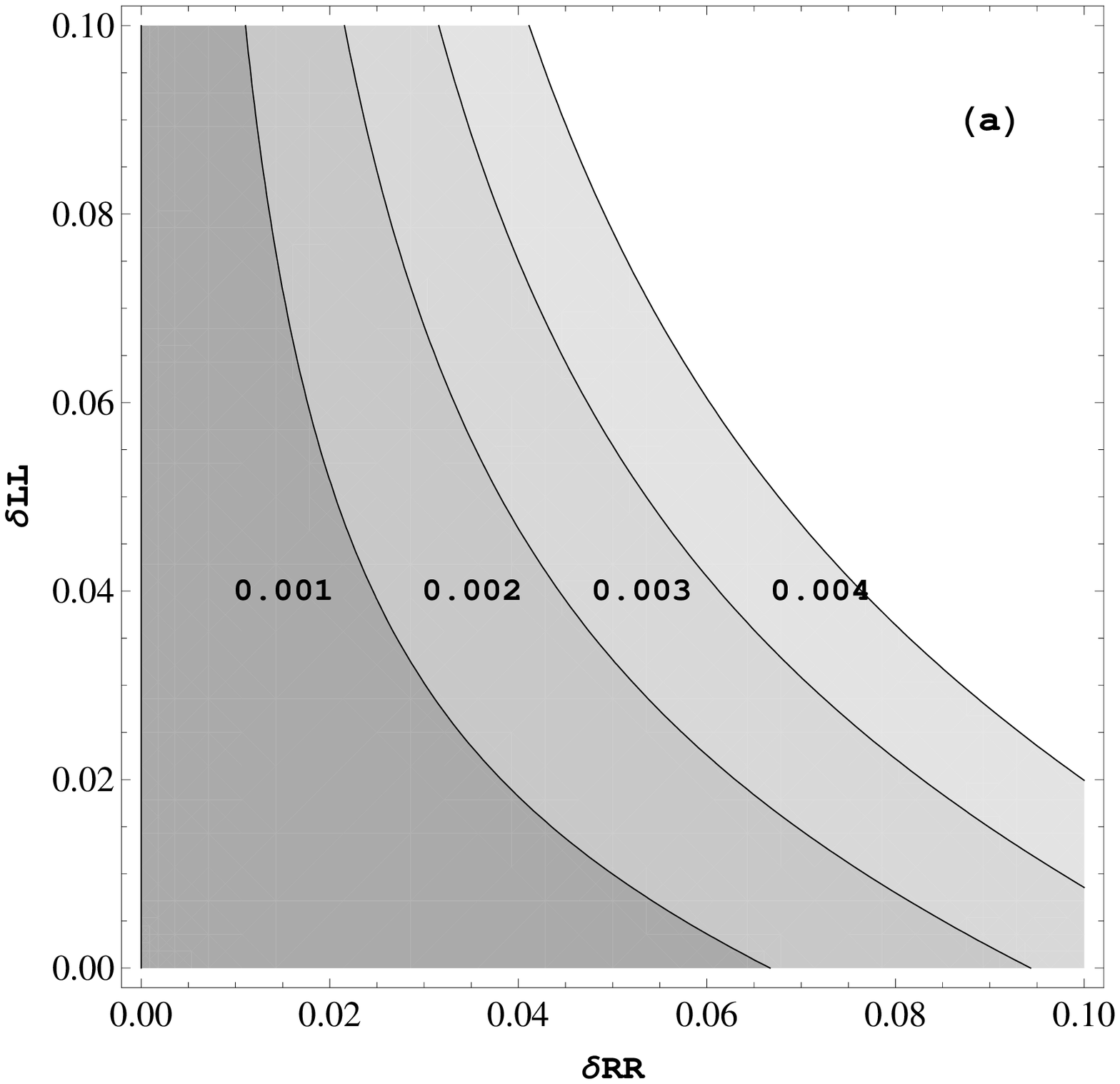} 
\includegraphics*[height=7.5cm]{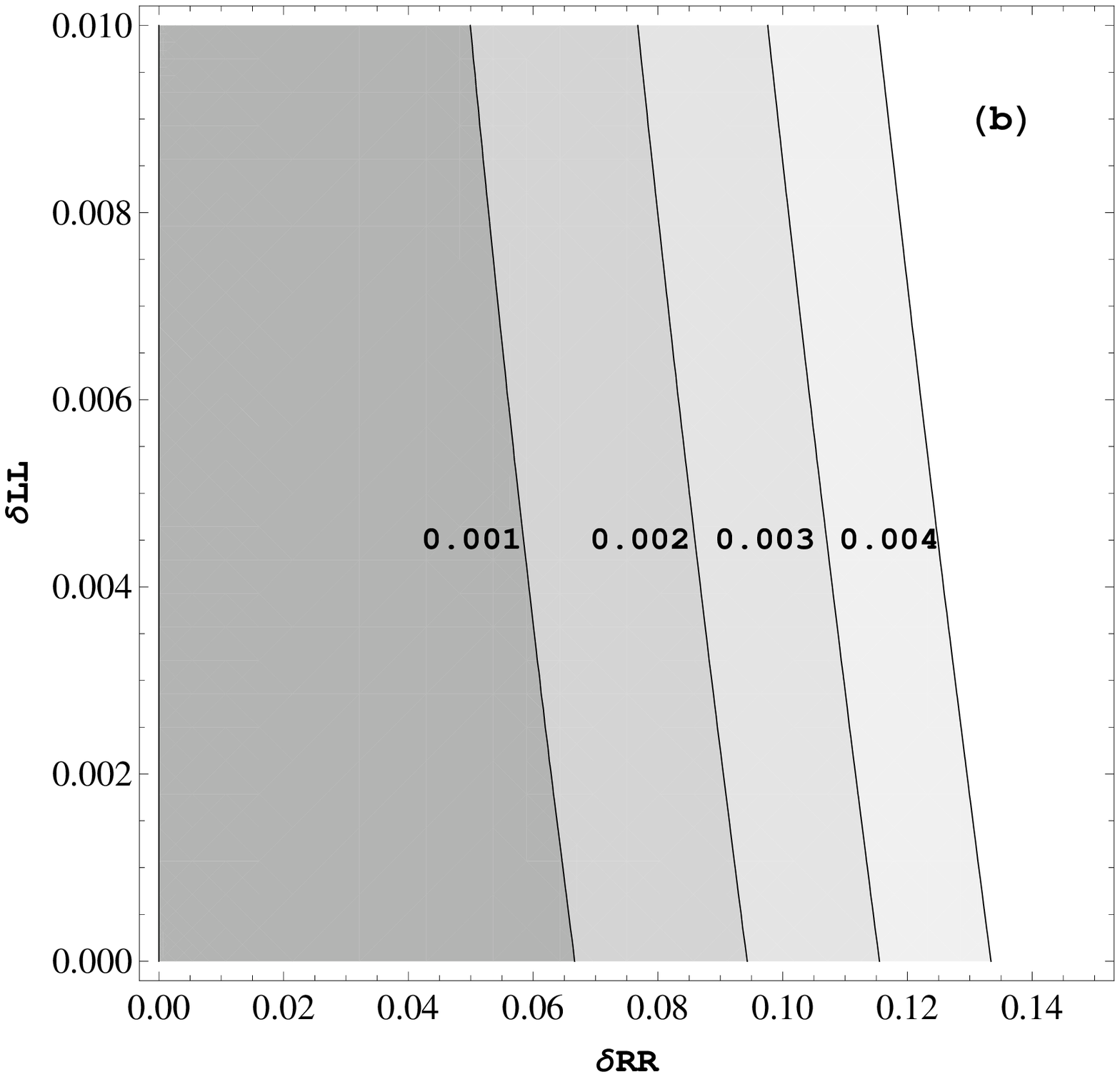} 
   \caption{ \it
Contour plots of the Lepton-Flavour-Violating (LFV) correction to the Standard
Model Value of $R_K$ (denoted by $\delta R_K$) as a function of the 
soft term mixing parameters, for $\tan\beta$ = 50.
We assume for illustration $M_H$ = 180~GeV, $M_1$ = 190~GeV, $\mu$ = 650~GeV,
$m_L$ = 300~GeV and $m_R$ = 200~GeV. 
In (a) the left and right slepton mixing are comparable, while
in (b) the right-handed slepton mixing dominates.}
 \label{omega}
\end{figure}

Before passing to the details of the calculation of the $\delta^{ij}_{XX}$
in GUT scenarios,  we  give
a feeling for the magnitudes of $\delta_{RR}$ and 
$\delta_{LL}$ required to see a signal in
$K\rightarrow \ell\nu$ decays, for realistic points in the supersymmetric
parameter space. This is done in  Fig.~\ref{omega}, which shows 
contour plots of the calculated 
deviation from universality in $R_K$, as functions of the $\delta^{ij}_{LL,RR}$.
Contour plot (a), on the left side,
indicates that if $\delta_{LL}$ and $\delta_{RR}$ were to be 
comparable, and the NA62 experiment
reaches the expected sensitivity of 0.003, it would be possible to observe
non-universality for slepton mixing parameters $\delta = {\cal O}(0.04-0.05)$, 
for a feasible set of parameters with a light Higgs boson and a light right-handed 
third-generation slepton mass. 
However, from the RGE running  one would naively expect that the left-slepton mixing
would be larger, and simultaneous mixing in the
$LL$ and $RR$ channels would tend to generate unacceptably large
flavour violation in channels that are strongly constrained
(particularly $\mu \to e \gamma$, which must be kept
under control in any LFV SUSY model). This would imply that for
non-negligible non-universality in kaon decays $\delta_{LL}$ would have to be
small. This would then correspond to solutions with a dominant
right-handed slepton mixing, as in the contour plot (b) on the right.

We examine in subsequent sections the magnitudes of right-handed slepton mixing
that arise in various theoretical scenarios.
We first discuss briefly the non-universal corrections to the soft sfermion masses that are induced
in the presence of  non-zero $A$-terms, by RGE effects between $M_{GUT}$ and low energies
in sample seesaw neutrino-mass models for the complete $3 \times 3$ mixing.
In this case the RR mixing is too small for any observable effect.
Subsequently, we consider the possible effects of RGE running above the GUT scale, 
where the overall right-handed mixing may be amplified in some GUT scenarios.

\section{RGE effects below and above the GUT Scale}

In supersymmetric seesaw models of neutrino masses, the RGEs between the GUT scale
and the heavy singlet neutrino mass scale generate non-universalities in the soft
supersymmetry-breaking scalar masses. These may have implications for
rare kaon decays that involve two charged leptons, as well
as charged leptons and pions, as have been studied in~\cite{Kaons-John}. In that work
only the dominant left-handed slepton mixing was considered, neglecting the
subdominant  mixing in the right-handed slepton sector.

To proceed, we assume universal initial conditions for the soft terms at the GUT scale, and
consider the RGEs including neutrino Yukawa couplings. We also assume 
a single common mass scale $M_N$ for the heavy singlet neutrinos
(which may easily be modified, see~\cite{JM}).
Then, in the leading-logarithmic approximation the RGE-induced soft supersymmetry-breaking
terms are given by
\bea
\left(\delta \tilde{m}_{\tilde L}^2\right)_{ij} & \approx & 
-\frac{1}{8\pi^2}(3m_0^2 + A_0^2)
({ Y_\nu^\dagger}{ Y_\nu}  +{ Y_e^\dagger}{ Y_e})_{ij}
\log\frac{M_{GUT}}{M_N}\ , 
\nn \\
\left(\delta \tilde{m}_{\tilde E}^2\right)_{ij} & \approx & 
-\frac{1}{4\pi^2}(3m_0^2 + A_0^2)
({ Y_e}{ Y_e^\dagger})_{ij}
\log\frac{M_{GUT}}{M_N}\ , 
\nn \\
(\delta {A_e})_{ij}& \approx &  
- \frac{1}{8\pi^2} A_0 Y_{e_i} (
3 { Y_e^\dagger}{Y_e}
+ { Y_\nu^\dagger}{Y_\nu})_{ij}
\log\frac{M_{GUT}}{M_N} .
\eea
The trilinear soft supersymmetry-breaking terms $A_{e,\nu}$ are assumed to be related by universal
factors $A_0$ to the corresponding Yukawa couplings $Y_{e,\nu}$. From this equation, it becomes
already clear that large values of $A_0$ could lead to enhanced RGE corrections to soft masses, a feature
that we will use in our considerations.

The above equations hold for fully universal initial conditions. However, 
well-motivated models with deviations from Higgs-scalar fermion universality,
as in \cite{noscH}, induce additional corrections
linked to  
$S = (M^2_{H_u} - M^2_{H_d}) + Tr_F (m^2_Q - 2 m^2_U + m^2_E +
m^2_D - m^2_L)$,
where the trace runs over flavours.

It is possible, even likely, that the GUT scale lies significantly below the
scale $M_{grav}$ at which gravitational effects can no longer be neglected.
In specific models, $M_{grav}$ might be identified with either the Planck mass
$M_P = 1.2 \times 10^{19}$~GeV or some lower string unification scale
$M_{string} \sim 10^{18}$~GeV. In general, the renormalization of couplings
at scales between $M_{grav}$ and $M_{GUT}$ may induce significant
flavour-violating effects, particularly in the $\delta^{ij}_{RR}$,
which can be calculated in any specific supersymmetric GUT.

The simplest example is provided by the minimal supersymmetric
SU(5) GUT, whose superpotential contains terms of the form $e ^c u^c \bar{H}$,
where the $\bar{H}$ is a colour-triplet Higgs field that is expected to have a 
mass $\sim M_{GUT}$. This gives rise to one-loop diagrams that
renormalize the right-handed slepton masses between $M_{GUT}$ and $M_{grav}$.
In the leading-logarithmic approximation, these take the form~\cite{HisNo}:
\begin{eqnarray}
(\delta \tilde m^2_{\tilde{E}})_{ij} 
&\simeq& - \frac{3}{8\pi^2}
  \lambda_{u_3}^2 V_{U}^{3i} V^{\ast 3j}_{U} 
 (3 m_0^2 +A_0^2)  \log  \frac{M_{\rm grav}}{M_{\rm GUT}} ,
 \label{VU}
\end{eqnarray}
for $i \ne j$, where $V_{U}$ denotes the 
mixing matrix in the corresponding couplings in the basis where
the $u$-quark and charged-lepton masses are diagonal. This is to be compared with
the corresponding corrections to left-handed slepton masses, which are proportional
to $V_{D}$, the Dirac neutrino mixing matrix in the basis where
the $d$-quark and charged-lepton masses are diagonal, and are given by
\begin{eqnarray}
(\delta \tilde m^2_{\tilde{L}})_{ij} 
&\simeq& - \frac{1}{8\pi^2} \left( 
  \lambda_{\nu_3}^2 V^{\ast 3i}_{D} V_{D}^{3j} 
 \log \frac{M_{\rm grav}}{M_{\nu_3}}
+ \lambda_{\nu_2}^2 V^{\ast 2i}_{D} V_{D}^{2j} 
 \log \frac{M_{\rm grav}}{M_{\nu_2}}
\right)  (3 m_0^2 +A^2_0)  .
\label{offdiagofL} 
\end{eqnarray}
Finally the leading-logarithmic renormalization of the $A_e$ terms is given by 
\begin{eqnarray}
\delta  A_e^{ij}&\simeq&
-\frac{3}{8\pi^2} A_0 \left( 
  \lambda_{e_i} V_{D}^{\ast 3i} V_{D}^{3j} \lambda_{\nu_3}^2 
       \log \frac{M_{\rm grav}}{M_{\nu_3}}  
+ \lambda_{e_i} V_{D}^{\ast 2i} V_{D}^{2j} \lambda_{\nu_2}^2 
       \log \frac{M_{\rm grav}}{M_{\nu_2}}  
\right. \nonumber \\
&&
\phantom{-\frac{3}{8\pi^2} A_0 }
\left. + 3 \lambda_{e_j} V_{U}^{\ast 3j} V_{U}^{3i} \lambda_{u_3}^2 
\log \frac{M_{\rm grav}}{M_{\rm GUT}} \right) .
\label{AVU}
\end{eqnarray}
One must appeal to a specific GUT model for the structures of the
mixing matrices $V_{U,D}$.

In the case of minimal supersymmetric SU(5), as already remarked, the $d$-quark 
mass matrix is the transpose of the charged-lepton mass matrix, and $V_D$ is
simply the unit matrix. On the other hand, $V_U$ is non-trivial, and related to the
familiar CKM matrix. We recall that in minimal SU(5) matter fields are
arranged in ${\mathbf {\bar 5}}$ ($(L,d^{c})_i$) and ${\mathbf{10}}$ supermultiplets
($(Q,u^{c},e^{c})_{i}$), the
$d$-quark and charged-lepton masses arise from ${\mathbf{10}}-{\mathbf {\bar 5}}-\bar{H}$
couplings $\lambda_{\bar5}$, and the $u$-quark masses arise from 
${\mathbf{10}}-{\mathbf{10}}-H$ couplings $\lambda_{10}$. The theory may be written in a
basis where the $\lambda_{\bar5}$ are diagonal, and hence also the
$d$-quark and charged-lepton masses. In this basis, the $d$-quark triplets in the
${\mathbf{10}}$ supermultiplets are rotated relative to the $u$-quark triplets and the
$u^c$ anti-triplets by the familiar CKM matrix $V_{CKM}$, and the
$u$-quark triplets and the $u^c$ antitriplets are related by a diagonal phase matrix $U$
with unit determinant~\cite{EGN}. It is clear from the forms of the equations (\ref{VU}, \ref{AVU})
that the phase matrix $U$ is irrelevant for our considerations in this paper,
though it might have played
a role in generating the baryon asymmetry of the universe~\cite{EGN2}.

In this simplest $SU(5)$ one has $m_b = m_\tau$ (a successful relation),
$m_s = m_\mu$ and $m_d = m_e$ (unsuccessful relations)~\cite{CEG}.
The latter predictions can be modified by taking into account possible
non-renormalizable fourth-order terms in the effective superpotential,
of the form $\bar{H}-{\mathbf {10}}-{\mathbf {24}}-{\mathbf {\bar 5}}$~\cite{EG}, which make
different contributions to the $d$-quark and charged-lepton mass matrices:
\[
\lambda ({\mathbf {10}}-{\mathbf {\bar 5}}-\bar{H}) + \lambda^{\prime} 
(\bar{H}-{\mathbf {10}}-{\mathbf {24}}-{\mathbf {\bar 5}}) 
\rightarrow 
\lambda \bar{v} (d d^c + e^c e) +
\lambda \bar{v} V (2 d d^c -3 e^c e),
\]
implying that, in the basis where $m_d$ is diagonal,
\beq
m_e = m_d^D - 5 \lambda^{\prime} \bar{v} V,
\eeq
where the matrix of couplings $\lambda^{\prime}$ is non-diagonal, in general.
Then, the diagonalization of $m_e^D = V_{eR} m_e V_{eL}^{+}$ gives
\beq
m_e^D = V_{eR} (m_d^D - 5 \lambda^{\prime} \bar{v} V) V_{eL}^{+} .
\eeq
Hence, in this modification of the simplest SU(5) model, the diagonalization of 
charged lepton mass matrix is not any more given by $V_{CKM}$ and the model
can realistically reproduce the observed phenomenology. 
In a similar manner~\cite{his2}, 
the colour-triplet-induced $e^c u^c$ mixing receives potentially large 
corrections for the first two generations. Parametrizing the non-renormalizable
correction to this mixing by $V_{uR},$ the RGE induced right-slepton mixing
is not given by $V_{CKM}$ as in the minimal model, but by the product 
\bea
V_R=V_{CKM} V^{+}_{uR}.
\label{VR}
\eea 
As $V_{uR}$ is not  constrained at present, 
we assume that all possible values of mixing angles 
parameterizing $V_R$ are allowed.

These non-renormalizable corrections also change 
the forms of the fermion mass matrices,
and hence the predictions of this type of flavour texture model within
minimal SU(5). For example, the predictions on new physics effects 
in $B_s-\bar B_s$ mixing~\cite{Hisano:2003bd} will be modified and the 
direct relation between the latter  and lepton flavour 
violating observables is lost.
Thus, these corrections would also affect the
renormalization between the GUT and heavy-neutrino mass scales.
These effects would also be important for the $\delta^{ij}_{LL}$, but
we do not consider them here.

Another GUT scenario is flipped SU(5), in which
the fields $Q_{i},d_{i}^{c}$ and $\nu_{i}^{c}$ 
of each family belong to a 
${\mathbf{10}}$ representation of SU(5), 
the $u_{i}^{c}$ and $L_{i}$ belong to 
${\mathbf{\bar{5}}}$  
representations, and the $e_{i}^{c}$ fields belong to
singlet representations of the group.

In this case, one would expect that large
right-handed slepton mixing could be accommodated more easily.
These particle assignments imply a symmetric down-quark mass matrix, 
and a charged-lepton mixing matrix that is not directly correlated with that
of the quarks, and the corresponding mixing angle and phase analysis has been
carried out in~\cite{ELNO}. However, the correlation between left-handed charged leptons and
right-handed $u$ quarks, as well as the direct link between the neutrino and $d$-quark
mass matrices, makes it hard to find a phenomenological model
with a U(1) flavour group that also accommodates
the solar and atmospheric neutrino data, without fine tuning of the
flavour charges~\cite{SU5-b, EGL}. In string-inspired versions of
flipped SU(5), natural solutions to the complete fermion data have been
found~\cite{ELLN}, but the large number 
of zero entries in the mass matrices imposed by string selection
rules leave room only for minimal flavour mixing, and we do not study them further here.

\section{SUSY (Flavour) Parameter Space and Numerical Examples}

In order to induce observable non-universality effects in $K \to \ell \nu$ decays
due to RGE effects below and above the GUT scale, while
respecting all available experimental constraints on flavour conserving and 
violating processes and cosmology, both the mass pattern of supersymmetry breaking terms
as well as their RGE induced flavour structure must be non-trivially constrained.
In this Section we first provide  an example of a supersymmetry breaking scenario
which satisfies all such constraints. Subsequently, we discuss the 
necessary flavour pattern of the soft supersymmetry breaking  parameters and we
find how such flavour structures can be RGE induced above and below the GUT scale.
We find that the necessary flavour structure for our scenario is very tightly constrained.

The study of models with a mass spectrum that could potentially lead to
large $K \to \ell \nu$ decays has been motivated 
for independent reasons. For instance, the WMAP benchmark 
scenarios with universal supersymmetry-breaking soft terms studied in~\cite{BDEGOP}
include some in the $\chi-{\tilde \tau_1}$ coannihilation region, which have light right-handed staus.
However, these scenarios generally predict high masses for the heavier
Higgs bosons, leading to a suppression of non-universality in 
$K\rightarrow \ell\nu$ decays. This then suggests moving to the study of models that
deviate from the minimal schemes, e.g., by
breaking the universality of the soft supersymmetry-breaking 
masses in the Higgs sector~\cite{Bench3}.
Indeed, soft universality in the Higgs sector
is not as well motivated as for the sfermion masses. Moreover,
large values for the $A$-terms would also allow smaller heavy Higgs 
masses~\cite{EHOW}.

When looking for input SUSY parameters at some high scale that are
consistent with supersymmetric Dark Matter and
with all experimental constraints, we consider the following 
region of the free parameters:
\bea
m_0 \ll M_\frac{1}{2} < |M_{H_d}|\approx |M_{H_u}|,\; A_{0}; \; \tan\beta >50, \; sign(\mu).
\label{param}
\eea  
This scenario resembles the so-called Higgs boson exempt no-scale supersymmetry breaking
scenario \cite{noscH}. In this scheme, 
all the RGE-corrected SUSY breaking masses at low scale
are large, thus explaining why no SUSY particles have been observed so far.
The Higgs mass parameters $|M_{H_d}|^2 \approx |M_{H_u}|^2<0$ are negative, 
triggering the electroweak symmetry breaking. 
However, the light charged Higgs mass (as well as the correct scale of the
electroweak symmetry breaking)   are obtained 
 due to the large cancellations between the RGE-corrected SUSY parameters, 
and are thus tuned. Since $m_0$ is smaller than all other 
 parameters of the model, RGE-induced LFV is generated by the large
 parameters $ M_{H_u},\; A_{0}$ via
 \begin{eqnarray}
 \left(\delta \tilde{m}_{\tilde L}^2\right)_{ij} & \approx & 
-\frac{1}{8\pi^2}(M_{H_u}^2 + A_0^2)
({ Y_\nu^\dagger}     \log\frac{M_{\rm grav}}{M_N}\   { Y_\nu}  )_{ij} , 
\label{rge1}
\\
(\delta \tilde m^2_{\tilde{E}})_{ij} 
&\simeq& - \frac{3}{8\pi^2}
  \lambda_{u_3}^2 V_{R}^{3i} V^{\ast 3j}_{R} 
 (M_{H_u}^2 + A_0^2)  \log  \frac{M_{\rm grav}}{M_{\rm GUT}} , 
 \label{rge2}
 \end{eqnarray}
where $V_R$ is the mixing matrix \rfn{VR} corrected by the higher-dimensional operators and
we have assumed that the top quark Yukawa coupling $ \lambda_{u_3}$ does not
receive large corrections.
Thus, large off-diagonal
 elements in both the left- and right-slepton mass matrices are to 
 be expected at low energies.

As a representative example, we take the parameter set appearing in 
Table \ref{paramT}, which results in a heavy sparticle spectrum, but a light
charged Higgs boson mass. The Higgs mass is fine-tuned, 
so small changes in the parameters would 
alter the charged 
Higgs boson mass significantly. Moreover, since
small changes in $M_\frac{1}{2}, \; |M_{H_d}|,\; |M_{H_u}|,\; A_{0}$
do not drastically alter the rest of the model parameters, 
in this scenario, the charged Higgs boson mass 
can essentially be considered as a free parameter.
For example, increasing (decreasing) the Higgs mass parameter $|M_{H_u}|$ 
by 5 GeV compared to the value in  Table \ref{paramT},  would result to
a charged Higgs boson of 235~GeV (159~GeV).
In the numerical study that follows, we perturb the parameters 
\rfn{param} around the values of Table \ref{paramT},
so that a light charged Higgs boson is obtained. 

\begin{table}[h]
\begin{center}
\begin{tabular}{||c|c||}
\hline \hline
Input Parameters  & Value   \\
\hline \hline 
$m_{1/2}$      & 1000   \\
$m_0$          & 200   \\
$\tan{\beta}$  & 50    \\ 
$|M_{H_u}|$ & 2550  \\
$|M_{H_d}|$ & 2500  \\
$A_0$         & 3000   \\
\hline \hline
(s)-particle masses  & Value  \\
\hline \hline
$M_1$ & 432\\
 $m_{\chi_1}$  & 425  \\
$\mu$ &  2394\\
$m_{h0}$ &  115\\
$M_{H^+}$ & 201 \\
$m_{e_L}$, $m_{\mu_L}$ & 681  \\
$m_{e_R}$, $m_{\mu_R}$ &  432 \\
$m_{\tau_1}$  &  505  \\
$m_{\tau_2}$ &   852  \\
\hline \hline
\end{tabular}
\end{center}
\caption{\it Sample supersymmetric particle spectrum that may lead to enhanced
non-universality in $K\rightarrow \ell\nu$ decays. The mass parameters are in GeV units.
}
\label{paramT}
\end{table}

We now turn to discuss the constraints on the flavour structure 
of the SUSY mass matrices. In general, as Section 3 indicates,
non-vanishing neutrino masses and large mixing in the neutrino sector 
imply large LFV effects in the LL slepton sector in SUSY seesaw models.
Moreover, the simultaneous presence of LL and RR slepton 
mixing in the large-$\tan\beta$
regime implies enormous enhancement of LFV decay rates. As the RR mixing is 
necessary for observable non-universality in the kaon decays under discussion,
we have to forbid any significant  mixing in the LL slepton sector. 
This is greatly facilitated by using the parameterization of neutrino seesaw parameters,
in terms of effective light neutrino observables and an auxiliary Hermitian
matrix $H$ \cite{Parametrisation} that can be related directly to low-energy
observables, including the processes that violate lepton number.
Indeed, the Hermitian matrix $H$ in the leading-logarithmic approximation can be regarded as 
the FLV mixing in the LL slepton sector, and is given by
\beq
H_{ij} = \sum (Y_{\nu}^*)_{ki}(Y_{\nu})_{kj}
\log\frac{M_{GUT}}{M_{N_k}} .
\eeq 
Observable neutrino masses and mixing can be obtained for 
\bea
H&=&
\left(
\begin{array}{ccc}
H_{11} & 0 & 0 \\
0 & H_{22} & 0 \\
0 & 0 & H_{33} \\
\end{array}
\right) ,
\label{H}
\eea
which minimizes at leading-logarithmic level all flavour mixings in the left slepton sector.
\eq{H} also implies that the CP violation in the neutrino sector 
is entirely linked to leptonic CP violation in the 
light neutrino sector, i.e., to the Dirac phase $\delta$ and to the 
two Majorana phases $\beta_{1,2}$ of the light neutrino mass matrix.
These phases give rise to CP violation consistent with leptogenesis~\cite{Ellis:2002xg}, 
as well as to electric dipole moments of charged leptons~\cite{JM,Masina:2003wt}.
Consequently, in this scenario, high-energy CP violation in 
$N_i$ decays can, in principle,
be tested through low-energy measurements.

As already discussed, in contrast to the left-slepton sector, 
large flavour mixings in the right-slepton sector must exist in order to 
generate observable non-universality in the $K\rightarrow \ell\nu$  decays.
Specifically, as discussed earlier, the mixing must be large 
in the  $\tau -e$ sector; such a mixing could  be 
induced due to the RGE running above the GUT scale. 
However if, in addition, considerable mixing exists in the $\mu-e$ or $\mu-\tau$ sectors,
the stringent experimental bounds from $\mu\to e\gamma$  decays 
would rule out the scenario. 
Thus the  phenomenological requirements are such that  
only $(1 \leftrightarrow 3)$ LFV mixing is allowed in the RR
sector.   The above considerations indicate that the 
SU(5) GUT model must be  non-minimal and 
 fine-tuned in the flavour sector above the GUT scale. 
In practice this implies that the non-renormalizable corrections  \cite{his2}
to the coloured triplet Higgs Yukawa couplings must be such  
that the corrected mixing matrix \rfn{VR} is, in the standard parameterization,
 described with the mixing angles 
$\theta_{12}^R=\theta_{23}^R=0,$ $\theta_{13}^R\neq 0.$ 
We do not speculate on the origin of such a flavour
pattern, we just comment that such a model is consistent with phenomenology and
allowed by model building \cite{his2}. 

In conclusion, the flavour constraints on
the non-universality parameter $\Delta^{31}_R$, the decay
$BR(\tau\to e\gamma)$ and the EDM of the electron can
depend only on \eq{H} which controls the heavy neutrino parameters and on the
$(1 \leftrightarrow 3)$ mixing
in the RR slepton sector. Thus the flavour structure is essentially fixed, implying 
particularly strong correlations between the relevant observables.

\begin{figure}[!h]
\begin{center}
\includegraphics*[height=7.5cm]{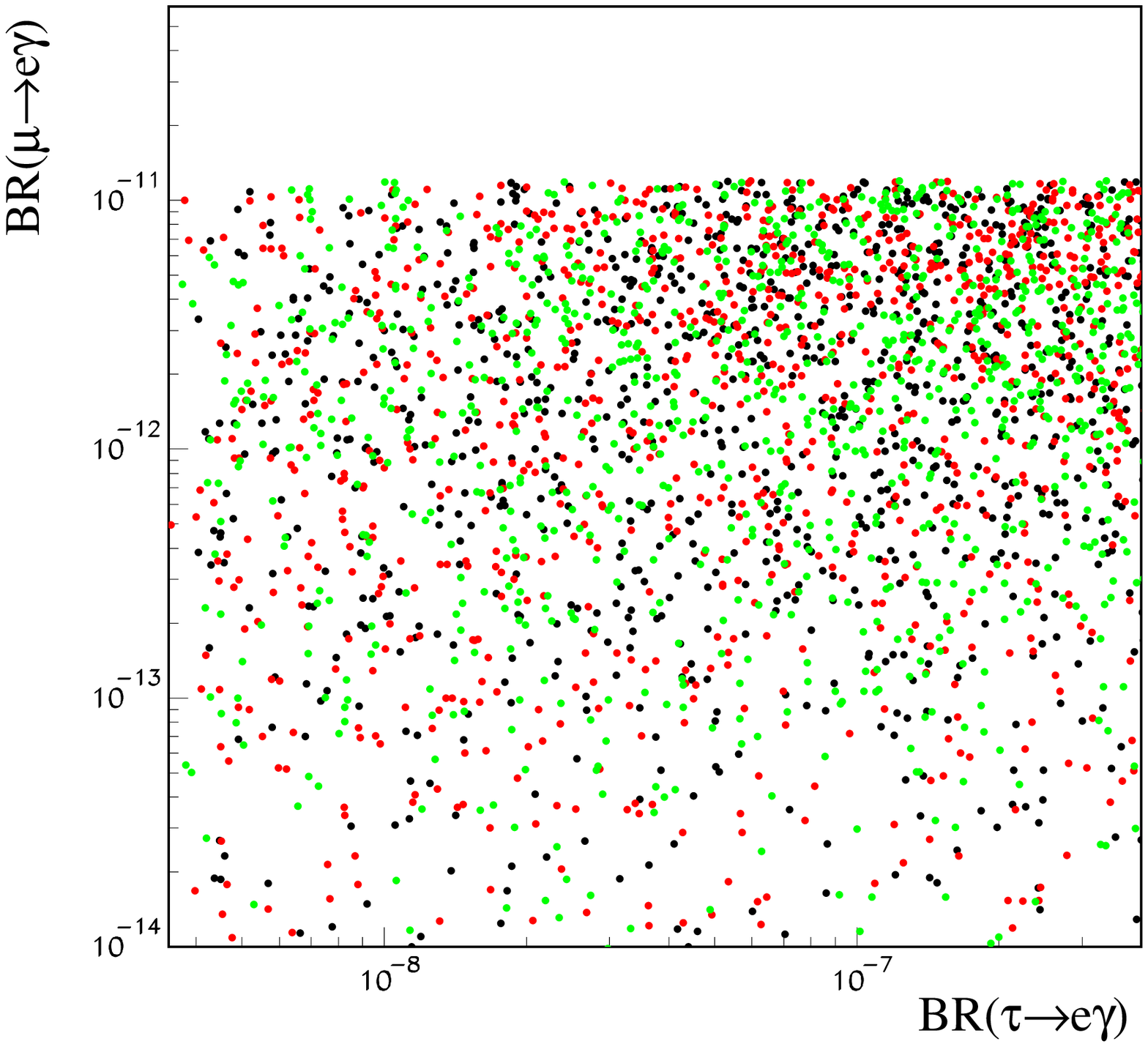} 
\includegraphics*[height=7.5cm]{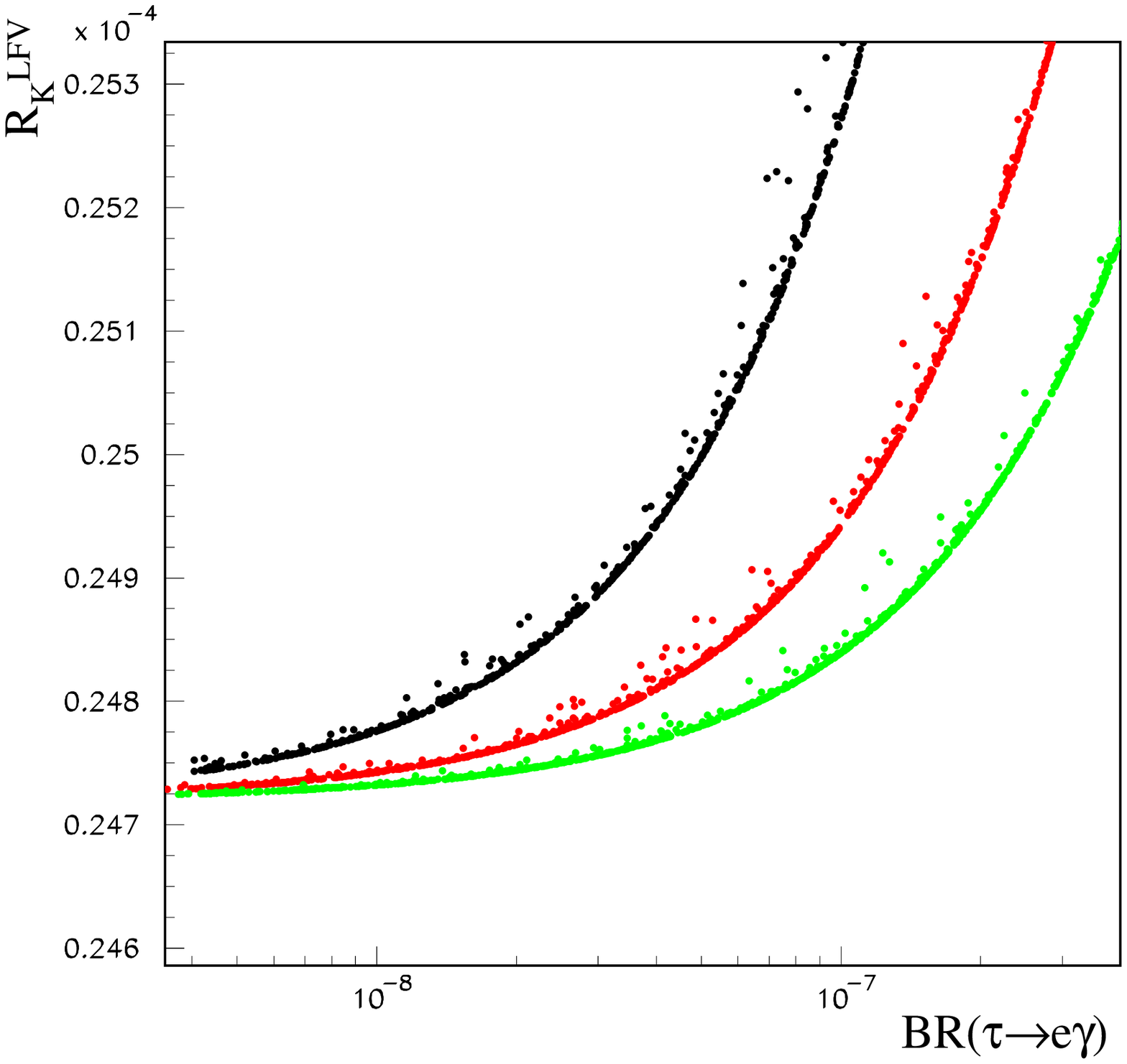} 
   \caption{ \it
Correlations  of  $BR(\mu\to e\gamma)$ (left panel) and 
$R_K^{LFV}$  (right panel)  with  $BR(\tau\to e\gamma)$
for the SUSY points $\tan\beta=50,$ $M_\frac{1}{2}=1000$ GeV, $m_0=200$ GeV, $A_0=3000$ GeV,
$|M_{H_d}|=2500$ GeV,  and for three values of $|M_{H_u}|=2545$ GeV (upper band, black dots),  
$|M_{H_u}|=2550$ GeV (middle band, red dots),  $|M_{H_u}|=2555$ GeV (upper band, green dots),  
The remaining parameters are randomly generated.
 }
 \label{fig2} 
 \end{center}
\end{figure}

 Although the flavour construction presented above eliminates the 
 most constraining RGE-induced
 LFV decay $\mu\to e \gamma$ at the leading-logarithmic level, dangerous 
 $(1 \leftrightarrow 2)$  mixing
 appears beyond the leading-logarithmic approximation. Our first concern is
to check that our  numerical 
 calculations are consistent with all the present bounds on LFV decays. In the left panel 
 of Fig.~\ref{fig2} we present a scatter plot of the branching ratios for 
 $BR(\mu\to e \gamma)$ and $BR(\tau\to e \gamma)$ which are obtained for the SUSY
 point of Table \ref{paramT} by randomly 
 generating all the free neutrino seesaw parameters and the right-mixing 
angle $\theta_{13}^R$. Since we work in the large-$\tan\beta$ regime, 
the value of $BR(\mu\to e \gamma)$
 generated beyond the leading-logarithmic level can be as large as the present experimental bound.
  The decay  rate of $\tau\to e \gamma$ is directly controlled by $\theta_{13}^R$, which 
 constrains its value.
  There is no  correlation between  $BR(\mu\to e \gamma)$ and the
 other observables of the model and, thus, both $BR(\mu\to e \gamma)$
 and $BR(\tau\to e \gamma)$ can be suppressed
 below the present bound by our flavour construction.
 
 However,   the right panel of Fig.~\ref{fig2}, 
 in which we plot the non-universality parameter $R_K^{LFV}$ as a function of 
  $BR(\tau\to e \gamma)$ for three different charged Higgs boson masses, 
 indicates a strong correlation between these observables. 
 We conclude that non-universality of this magnitude is indeed observable 
 in the NA62 experiment. Detection of non-universality 
in $K\rightarrow \ell\nu$ decays would allow
estimating the LFV rates in the tau sector, provided that the charged Higgs boson mass
 is determined at the LHC.

\begin{figure}[!h]
\begin{center}
\includegraphics*[height=7.5cm]{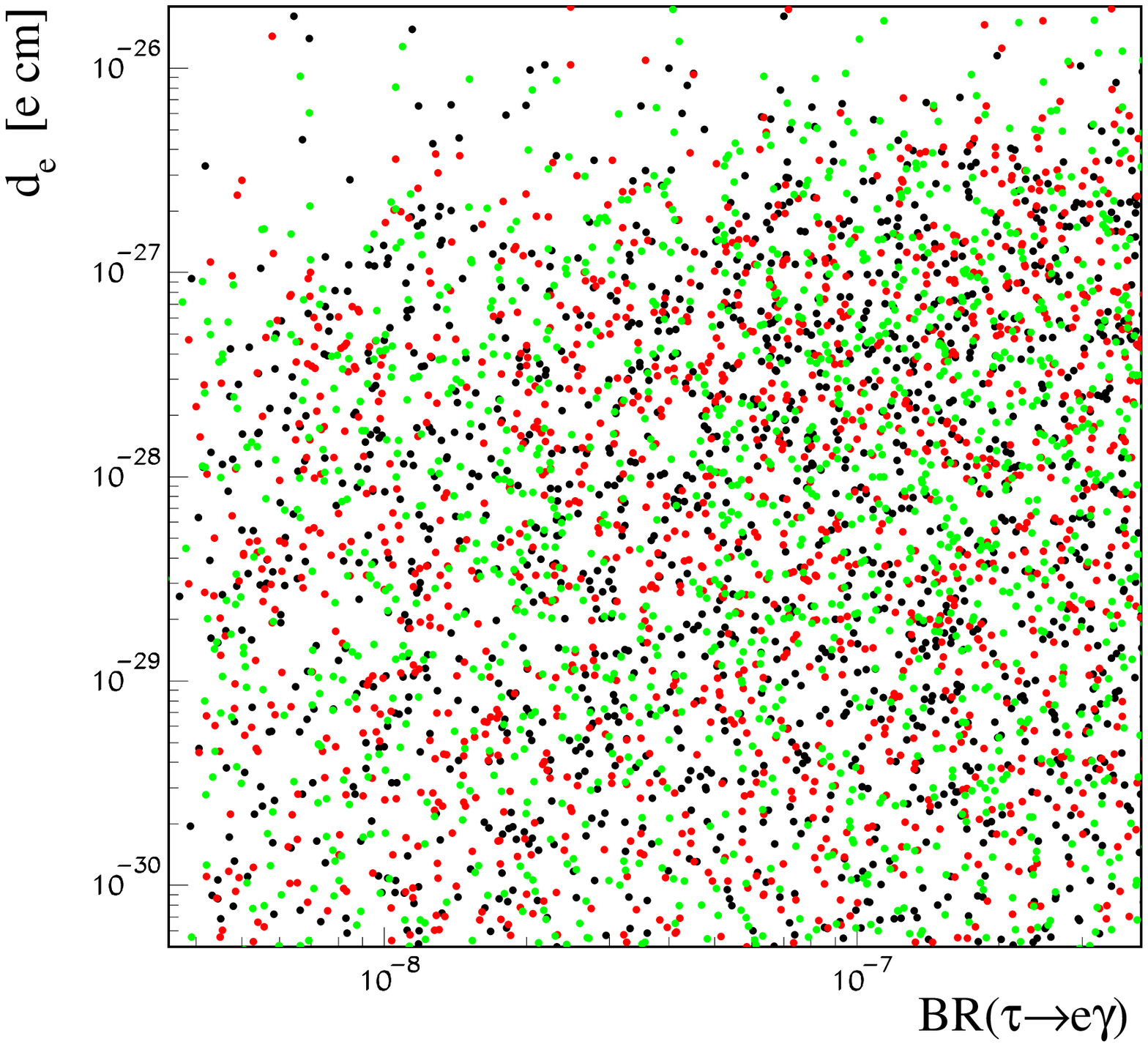} 
\includegraphics*[height=7.5cm]{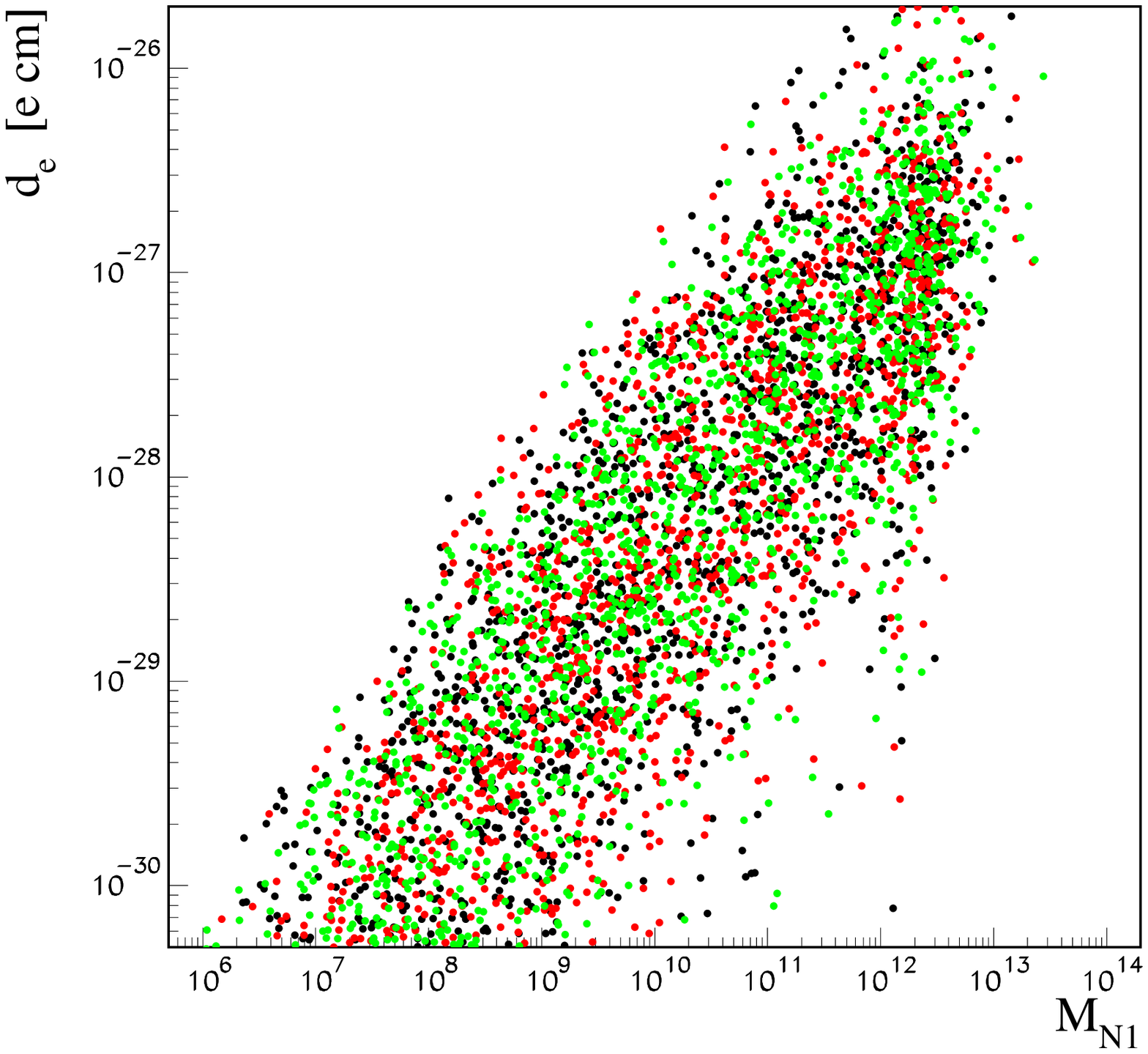} 
   \caption{ \it
Dependence of the electron electric dipole  moment
$d_e$ on $BR(\tau\to e\gamma)$ (left panel) and on the 
lightest neutrino mass $M_{N_1}$  (right panel).
The parameters and the colour code are as in Fig.~\ref{fig2}. }
 \label{fig3}
 \end{center}
\end{figure}

We now recall a couple of well-known generic problems in SUSY:
the supersymmetric CP problem and the
cosmological gravitino problem.
 In our scenario the parameterization \rfn{H} solves both of them 
provided $H_{11}\ll H_{22,33}.$  Indeed, the EDM of electron, $d_e,$ 
is proportional to $H_{11}$, and its smallness suppresses this EDM independently of the
phases arising in other sectors of the theory.
At the same time, in this parameterization, \eq{H} also determines the heavy neutrino mass
spectrum.
If $H_{11}\ll 1,$ the seesaw mechanism implies that the
$N_1$ mass has to be small, allowing a low reheating temperature  of the
Universe  and solving the gravitino problem.
Thus, if our construction is correct, 
there should be a correlation between the maximal $d_e$ and the lightest neutrino mass. 
In such a case, the standard Fukugita-Yanagida leptogenesis mechanism~\cite{Fukugita:1986hr} 
cannot provide the observed baryon asymmetry of the universe due to
too small  $M_{N_1}$ ~\cite{Davidson:2002qv}.
In our SUSY scenario, therefore, resonant leptogenesis or  
``soft leptogenesis''~\cite{D'Ambrosio:2003wy}  turn out  to be the 
favoured leptogenesis mechanisms.

We now study quantitatively the above qualitative statements. 
We first recall that in the minimal SUSY seesaw model (without right-slepton mixings induced above
the GUT scale)  one finds strong correlations between the generated baryon asymmetry, 
the RGE-induced electron
electric dipole moment  $d_e$ and $BR(\tau\to e\gamma)$~\cite{Ellis:2002xg}.
The maximally allowed values of $d_e$ are a few orders of magnitude below the present
experimental bound.
This correlation occurs because in the minimal SUSY seesaw model all these observables 
are generated by the dominant $(1 \leftrightarrow 3)$ mixing. In our scenario the 
$(1 \leftrightarrow 3)$ mixing
occurs in the right-slepton sector, and such a correlation is expected to be absent.
In the left panel of Fig.~\ref{fig3} we present a scatter plot of the values of   $d_e$ and 
$BR(\tau\to e\gamma)$ for the same parameters as in Fig.~\ref{fig2}.
In our scenario, the predicted values of  $d_e$ can easily exceed the present 
bound $d_e<1.6\cdot 10^{-27}$~e~cm and
the CP-violating and LFV observables are not correlated.
However, as explained above, there is a correlation between $d_e$ and $M_{N_1}$,
as can be seen in the right panel of Fig.~\ref{fig3}.
Indeed, for a fixed  $M_{N_1}$ there is an upper bound on the electron EDM.
Therefore our scenario relates the solution of the SUSY CP problem to the gravitino problem -
if $M_{N_1}$ is small enough to be generated thermally at reheating, 
the electron EDM  is suppressed. Thus, avoidance of the
gravitino problem in SUSY models could also explain why
$d_e$ has  not been observed so far.

\section{Conclusions}

In view of the  expected improvements in measurements of $K\rightarrow \ell\nu$
decays by the NA62 experiment, we have studied the expected 
violation of lepton universality in these decays, in supersymmetric models.
Unlike flavour-violating decays, which mainly probe the left-handed sector
of the theory, a violation of universality in $K\rightarrow \ell\nu$ originates directly from mixing
effects in the right-handed slepton sector. In this respect, it would
provide a unique probe into this aspect of 
supersymmetric flavour physics, particularly for large $\tan\beta$.

Unless universality in the scalar soft terms is violated,
$K\rightarrow \ell\nu$ decays can give observable rates only
in non-minimal grand unified models; this would occur through
a combination of RGE effects above the GUT scale and
higher-dimensional terms that enhance the mixing among the right-handed sleptons. 
Even in this case, we are limited to  very specific regions of the parameter space, 
with large $A$ terms and small Higgs masses. 
Moreover,  the very strong bounds from several flavour-violating processes
would require a significant suppression of left-handed slepton mixing, would 
further  limit the already constrained the supersymmetric parameter space,
and would imply fine-tuned solutions.

We find that, in the scenario under consideration, the flavour structure of the soft
supersymmetry breaking terms induced by RGE effects both below and above
the GUT scale is essentially fixed. This implies strong correlations between 
different lepton-flavour-violating
processes. In particular, should the NA62 experiment at CERN discover the 
non-universality effects,  observable rates for $\tau\to e\gamma$ can be predicted. 
At the same time, the electron EDM naturally exceeds the present experimental bound
unless the lightest heavy neutrino mass is sufficiently small,
as seen in Fig.~\ref{fig3}. In this scenario the solution to the 
supersymmetric gravitino problem is, due to the constrained flavour structure of 
the neutrino Yukawa couplings, related to the LFV observables and EDMs. 
In particular, the expected future experimental sensitivity to the electron EDM will put an 
upper limit to the lightest heavy neutrino mass and to solve (or rule out this solution)
to the gravitino problem.

In view of the above, one may hope for either of the following: \\
(i)  to see a deviation from lepton universality  in the near future, which would
imply that we must focus on a very constrained set of solutions in the SUSY parameter space;\\
(ii) to obtain further constraints on the model parameters and unknown
aspects of right-handed fermion and sfermion mixing.

\vspace*{0.2 cm}
{\bf Acknowledgements} 
We thank P. Paradisi for useful communication and discussions.
S. Lola and M. Raidal would like to thank the CERN Theory Division, where
a significant amount of this research has been performed. 
The research of S. Lola is funded in part by the FP6  Marie Curie Excellence
Grant MEXT-CT-2004-014297 and that of M. Raidal  by the ESF grant No. 6140.
The work of J. Ellis and S. Lola was supported in part by the European Union through the Marie
Curie Research and Training Network UniverseNet (MRTN-CT-2006-035863).

\end{document}